%% file: main.tex
\documentclass[11pt]{article}
\usepackage{amsmath}
\usepackage{amssymb}
\usepackage{amsthm}
\usepackage{todonotes}
\usepackage{tikz}
\usepackage{pgfplots}
\usepackage{graphicx}
\usepackage{epstopdf}
\usepackage[labelfont=bf,font=footnotesize]{caption}
\usepackage[labelfont=bf,font=footnotesize]{subcaption}
\usepackage{xcolor}
\usepackage{algpseudocode}
\usepackage{algorithm}
\usepackage{bbm}
\usepackage{makecell}
\usepackage{authblk}
\usepackage{fullpage,etoolbox}
\usepackage[colorlinks=true,citecolor=blue,linkcolor=blue,urlcolor=blue]{hyperref}
\usepackage[shortlabels]{enumitem}
\usepackage[round,sort,compress]{natbib}
\usepackage{listings}

\input{commands.tex}

\title{Communication Topology Co-Design in Graph Recurrent Neural Network Based Distributed Control}
\author{Fengjun Yang$^{\dag}$ and Nikolai Matni$^{\ast}$
\thanks{$^\dag$F. Yang is with the Dept. of Computer and Information Sciences, University of Pennsylvania, Philadelphia, PA. $^\ast$N. Matni is with the Dept. of Electrical and Systems Engineering, University of Pennsylvania, Philadelphia, PA. }%
\thanks{N. Matni is generously supported by NSF award CPS-2038873 and NSF CAREER award ECCS-2045834.}
}

\begin{document}
\maketitle

\begin{abstract}
When designing large-scale distributed controllers, the information sharing constraints between sub-controllers, as defined by a communication topology interconnecting them, are as important as the controller itself.  Controllers implemented using dense topologies typically outperform those implemented using sparse topologies, but it is also desirable to minimize the cost of controller deployment. Motivated by the above, we introduce a compact but expressive graph recurrent neural network (GRNN) parameterization of distributed controllers that is well suited for distributed controller and communication topology co-design.  Our proposed parameterization enjoys a local and distributed architecture, similar to previous Graph Neural Network (GNN)-based parameterizations, while further naturally allowing for joint optimization of the distributed controller and communication topology needed to implement it.  We show that the distributed controller/communication topology co-design task can be posed as an $\ell_1$-regularized empirical risk minimization problem that can be efficiently solved using stochastic gradient methods.  We run extensive simulations to study the performance of GRNN-based distributed controllers and show that (a) they achieve performance comparable to GNN-based controllers while having fewer free parameters, and (b) our method allows for performance/communication density tradeoff curves to be efficiently approximated.
\end{abstract}

\input{tex/1-intro}
\input{tex/2-Prelim}
\input{tex/3-Formulation}
\input{tex/4-Experiments}
\input{tex/5-conclusion}

\bibliographystyle{unsrtnat}
\bibliography{references}

\end{document}

%% file: commands.tex
\newcommand{\R}{\ensuremath{\mathbb{R}}}

\newcommand{\N}{\ensuremath{\mathbb{N}}}

\numberwithin{equation}{section}

\renewcommand{\geq}{\geqslant}

\renewcommand{\leq}{\leqslant}

\renewcommand{\succeq}{\succcurlyeq}

\newcommand{\minimize}{\mathrm{minimize}}

\newcommand{\vertexset}{\mathcal{V}}
\newcommand{\edgeset}{\mathcal{E}}
\newcommand{\dist}{\mathbf{dist}}
\newcommand{\infoset}{\mathcal{J}}
\newcommand{\totalinfoset}{\mathcal{I}}
\newcommand{\incoming}{\mathbf{in}}
\newcommand{\vectorize}[1]{\textrm{vec}(#1)}
\def\rmU{{\mathbf{U}}}
\def\rmX{{\mathbf{X}}}
\def\rmZ{{\mathbf{Z}}}

%% file: tex/1-intro.tex
\section{Introduction}
\label{sec: introduction}

Future large-scale autonomous systems, such as intelligent transportation, power, and communication networks, will be composed of dynamically interconnected subsystems and data-driven decision makers.  When designing the distributed control policies for such systems, the controller architectures, as defined by the actuation, sensing, and communication topologies of the controller, can no longer be taken for granted, as controllers with denser architectures will typically outperform controllers implemented using less resources. However, it is also desirable to minimize the cost of deploying controllers, leading to a non-trivial tradeoff between controller performance and architectural complexity.

A rich body of work exists addressing the controller architecture co-design problem in the context of (distributed) linear optimal control.  When no communication constraints are imposed on the controller, i.e., when the control problem is centralized, it has been shown that several metrics defined in terms of the system controllability/observability Gramians are (sub)modular.  This allows for efficient algorithms to be deployed for actuator/sensor selection that enjoy approximation guarantees, see for example \citet{pequito2015minimum,tzoumas2016sensor,summers2014optimal,summers2015submodularity} and references therein.  When communication constraints are present, augmenting the distributed control problem with a suitable regularizer to encourage simplicity in the controller architecture has proven to be successful.  For example, $\ell_1$ regularization is used to co-design sparse static state feedback gains in \citet{lin2013design} and synchronization topologies in \citet{fardad2014design}; group norm penalties are used to co-design actuation/sensing schemes in \citet{matni2016regularization,dhingra2014admm}; and a specialized atomic norm \citep{chandrasekaran2012convex} is used to design communication delay constraints that are well-suited to $\mathcal{H}_2$ distributed optimal control in \citet{matni2015communication}.  

In this paper, we assume that the actuation and sensing architecture is fixed, and focus on co-designing the communication topology of the distributed controller.  In particular, we model a distributed controller as a collection of sub-controllers, each equipped with a fixed set of actuators and sensors, that exchange
their respective measurements with each other subject to communication delays imposed by the to-be-designed
underlying communication topology.  In general, one can select communication topologies that range in density from completely decentralized, i.e., that allow no communication between sub-controllers, to ``virtually centralized,'' i.e, that allow instantaneous communication between all sub-controllers.  

However, there are now well understood limits on the simplicity of the designed communication topology \citep{rotkowitz2005characterization,rotkowitz2010convexity,lessard2011quadratic} if we wish for the resulting distributed control synthesis task to be convex.  In particular a sufficient, and under mild assumptions necessary, condition for a distributed linear optimal controller to be specified by the solution to a convex optimization problem\footnote{For a more detailed overview of the relationship between information exchange constraints and the convexity of distributed optimal control problems, we refer the reader to \cite{bamieh2005convex,mahajan2012information} and the references therein.}  is that the communication delays between sub-controllers allow for information to be exchanged between them at least as quickly as their control actions propagate through the plant \citep{rotkowitz2010convexity}.  This has limited the subset of the Pareto frontier that could be explored in previous work to communication topologies that are at least as dense as the topology of the underlying plant (see for example, the base QI communication graph in \citet[\S4.2]{matni2015communication}).

Motivated by these limitations on tractability and topology sparsity, this paper proposes an alternative approach.  We draw upon the recent successes of \emph{graph neural network} (GNN)-based controllers in the context of distributed control \citep{tolstaya2020learning,gama2020graph,khan2020graph,gama2021distributed}, and propose a novel distributed controller and communication topology co-design procedure.  We propose a compact \emph{graph recurrent neural network} (GRNN) parameterization for distributed controllers, and show that the control parameters and communication topology, as encoded by the \emph{graph shift operator} defining the GRNN-based distributed controller, can be jointly optimized via an $\ell_1$-regularized empirical risk minimization problem.  Finally, through extensive empirical evaluations, we show that  (a) distributed controllers based on our GRNN parameterization can achieve performance comparable to previously proposed GNN-based parameterizations while having fewer free parameters, and (b) our method allows for performance/communication density tradeoff curves to be efficiently approximated, even in the extremely sparse communication topology regime.



%% file: tex/2-Prelim.tex
\section{Preliminaries}
\label{sec: preliminary}
\subsection{Distributed Optimal Control}
Consider a system composed of $N$ subsystems, with the $i$-th subsystem described by a state vector $x_{i}(t) \in \R^p$ and a control input $u_{i}(t) \in \R^q$. We model inter-subsystem interactions through a time-invariant unweighted digraph $\mathcal{G}_{dyn}=(\vertexset, \edgeset_{dyn})$. Each element in the vertex set $\vertexset=\{v_1, v_2, ..., v_N\}$ corresponds to a subsystem. The edge set $\edgeset_{dyn} \subset \vertexset \times \vertexset$ contains an element $(i,j) \in \edgeset_{dyn}$ only when subsystem $i$ directly influences the states of subsystem $j$ through its states or control inputs. We assume linear time-invariant dynamics, allowing us to write the dynamics at subsystem $i$ as
\begin{equation}
    x_{i}(t+1) = \sum_{j:(j,i)\in \edgeset_{dyn}} A_{ij}x_{j}(t) + B_{ij}u_{j}(t),
    \label{eq: subsystem linear dynamics}
\end{equation}
for suitable matrices describing local interactions between subsystems. 
Equivalently, the dynamics of the full system can be expressed in terms of the joint state $x(t):=(x_i(t))_{i=1}^N$ and control $u(t):=(u_i(t))_{i=1}^N$ vectors
as
\begin{equation}
    x(t+1) = Ax(t) + Bu(t),
    \label{eq: joint linear dynamics}
\end{equation}
where matrices $A$ and $B$ are constructed such that the full dynamics (\ref{eq: joint linear dynamics}) are consistent with the subsystem dynamics in (\ref{eq: subsystem linear dynamics}). Note that the full system matrices $(A,B)$ are sparse when the subsystems are sparsely interconnected, i.e., when the graph $\mathcal{G}_{dyn}$ is sparse.

Our goal is to efficiently explore the space of distributed controllers composed of sub-controllers that have access to a combination of instantaneous local information and delayed global information. Formally, the sub-controller for subsystem $i$ at time $t$ can directly observe a local information set $\infoset_{i,t}$, which might fail to capture the full state of system (\ref{eq: joint linear dynamics}). For example, $\infoset_{i,t}$ might only include the state of subsystem $i$, i.e. $x_i(t)$, but not those of other subsystems. We assume that local information can be exchanged between subsystems, subject to a delay defined by a communication topology, modeled as a time-invariant unweighted digraph $\mathcal{G}_{c}=(\vertexset, \edgeset_{c})$ (note that the communication topology $\mathcal{G}_c$ need not equal the interaction topology $\mathcal{G}_{dyn}$). If $(i,j) \in \edgeset_{c}$, subsystem $i$ can share its local information set $\infoset_{i,t}$ with subsystem $j$, with a $1$-step delay, i.e., subsystem $j$ gains access to $\infoset_{i,t}$ at time $t+1$. Subsystem $j$ can then relay this information to its neighbors in $\mathcal{G}_{c}$ with a further time delay.

Denoting the directed distance from vertex $i$ to vertex $j$ in a digraph $\mathcal{G}$ as $\dist^{\mathcal{G}}(i\rightarrow j)$, and vertex $i$'s $d$-hop incoming neighbors as $\incoming^{\mathcal{G}}_i(d):=\left\{ v_{j}\;|\; \dist^{\mathcal{G}}(v_{j} \rightarrow v_{i}) \leq d \in \mathbb{N} \right\}$, the information sharing constraints imposed by the communication topology $\mathcal{G}_c$  mean that at time $t$, the total information available to subsystem $i$ is given by
\begin{equation}
    \label{eq:total_info}
    \totalinfoset_{i,t}(\mathcal{G}_c) := \underbrace{\infoset_{i,t}}_{\textrm{local information}} \cup 
    \underbrace{\left( 
        \bigcup_{d=1}^t \infoset_{j \in \incoming^{\mathcal{G}_{c}}_{i}(d), t-d} \right)}_{\textrm{delayed global information}}.
\end{equation}
This then restricts the action taken by subsystem $i$ at time $t$ to be of the form
\begin{equation}
    u_i(t) = \gamma_{i,t} \left(\totalinfoset_{i,t}(\mathcal{G}_c)\right),
    \label{eq: info constraint}
\end{equation}
for $\gamma_{i,t}$ a suitable map from the total information set $\totalinfoset_{i,t}(\mathcal{G}_c)$ to control actions.  This formulation also makes clear that by varying the density of the communication topology defined by $\mathcal{G}_c$, we can vary the total amount of information available to each subsystem $i$.

In the following, we summarize recent results showing that GNNs offer a natural means of parameterizing such controllers, and introduce our proposed parameterization for control and communication delay co-design.

\subsection{Graph Neural Networks for Distributed Control}
GNNs are a class of neural networks, built around the notion of a graph filter, that are well-suited for processing and representing data with graph structure. They enjoy many additional desirable properties, such as being permutation equivariant and Lipschitz continuous to changes in the network \citep{gama2020stability,gama2021distributed}, making them natural candidates for distributed controllers.  Following conventions in the GNN literature, we rewrite the joint state $x(t) \in \R^{Np}$ and control vectors $u(t) \in \R^{Nq}$ as
\begin{equation}
    \rmX(t) = 
          \begin{bmatrix}
           x_{1}^{T}(t) \\
           \vdots \\
           x_{N}^{T}(t)
          \end{bmatrix}\in \R^{N \times p}
    , \rmU(t) = 
          \begin{bmatrix}
           u_{1}^{T}(t) \\
           \vdots \\
           u_{N}^{T}(t)
          \end{bmatrix}\in \R^{N \times q},
    \label{eq: stacked-states}
\end{equation}
as this representation more naturally allows for the required graph convolutions to be defined.  In what follows, we tailor our discussion to using GNNs in the context of distributed control.

\subsubsection{Graph Convolutions}
The notation in (\ref{eq: stacked-states}) allows one to express linear information exchanges as a matrix multiplication while enforcing the communication topology as a constraint on the multiplicand. Specifically, consider a matrix $S \in \R ^{N \times N}$ which satisfies the property that $S_{ij} \neq 0$ only if $(j,i) \in \edgeset_c$, and define the matrix $U \in \R^{N \times p}$ as $U=S\rmX$. Then,
\begin{equation}
    U_{ij}
    = \sum_{k=1}^{N} S_{ik} \rmX_{kj}
    = \sum_{k: (k,i) \in \edgeset_c} S_{ik} \rmX_{kj}
    = \sum_{k: v_{k} \in \incoming^{\mathcal{G}_c}_{i}(1)} S_{ik} \rmX_{kj},
    \label{eq: GSO multiplication}
\end{equation}
where the first equality follows from the definition of matrix multiplication, the second equality follows the definition of nonzero elements of $S$, and the third from the definition of the vertex $i$'s $1$-hop incoming neighbors.  This latter definition emphasizes that the $i$-th row of $U$ can be computed by subsystem $i$ using information received from $1$-hop incoming neighbors in $\mathcal{G}_c$, i.e., only the $\rmX_{kj}$ in (\ref{eq: GSO multiplication}) such that $v_k \in \incoming^{\mathcal{G}_c}_{i}(1)$ are needed.

In general, left multiplying a joint state or input vector by the matrix $S$ corresponds to communication and aggregation of data between $1$-hop neighbors, as defined by the communication topology $\mathcal{G}_c$. By associating a unit delay with such a communication and aggregation step, we can model the \emph{spatiotemporal} propagation of information across a graph, making this operator a natural tool for encoding information sharing constraints in a distributed control setting.  We note that such an operation is analogous to the time-shift operator in classical signal processing, and is thus referred to as the \textit{graph shift operator}. Examples of valid graph shift operators on a graph $\mathcal{G}$ include its adjacency matrix and its Laplacian. For the rest of the paper, we denote the set of valid graph shift operators on graph $\mathcal{G}$ as $\mathcal{S}_{\mathcal{G}} \subset \R^{N \times N}$.

Graph convolution extends the graph shift operation by aggregating information over multiple time steps. Consider a graph $\mathcal{G}$ and a corresponding graph shift operator $S \in \mathcal{S}_{\mathcal{G}}$. We can define a \emph{linear distributed controller} $\rmU(t)$ in terms of the graph convolution of the joint state $\rmX(t)$ over $K$ steps specified by the set of weights $H=\{H_k\}_{k=0}^{K-1}, H_k \in \R^{p \times q}$ as
\begin{equation}
    \rmU(t) = H_{\mathcal{G}} \circledast \rmX(t) = \sum_{k=0}^{K-1} S^k \rmX(t-k) H_k,
    \label{eq: linear graph filter}
\end{equation}
where each row of resulting control signal $\rmU(t) \in \R^{N \times q}$ can be computed at subsystem $i$ using $k$-step delayed information received from subsystems $j$ such that $\dist_{\mathcal{G}}(j \rightarrow i) = k \leq K$. To see this, we first note that, by the properties of graph shift operators, $S^{k}_{ij} \neq 0$ only if $v_{j} \in \incoming_{i}^{\mathcal{G}}(k)$, i.e., node $j$ is a $k$-hop incoming neighbor of node $i$ in $\mathcal{G}$. Thus, for each $k \in \{0, ..., K-1\}$ in the sum, the $i$-th row of $\rmU(t)$ aggregates $k$-step delayed information from $k$-hop incoming neighbors of subsystem $i$, with information set given by $\infoset_{i,t} = \{ x_i(t) \}$.  We therefore have that the total information set needed to compute the $i$-th row of $\rmU(t)$ is given by 
\begin{align*}
\totalinfoset_{i,t}(\mathcal{G}_{c}) &= \{ x_i(t) \} \cup 
\left( 
        \bigcup_{k=1}^{K-1} \{ x_j(t-k) \, | \, j \in \incoming^{\mathcal{G}_{c}}_{i}(k) \} \right),
\end{align*}
i.e., the collection of $k$-step delayed states from $k$-hop neighbors, for $k=0,\dots, K-1$.
We highlight that the graph convolution encodes both spatial and temporal structure, as captured by the propagation of signals by repeated application of the graph shift and time delay operator within the convolution, making it well-suited for parameterizing controllers subject to information sharing constraints as in \eqref{eq: info constraint}.  Finally, we note that a filter $H_k$ is applied uniformly across all rows of its multiplicand. Thus, the result from right multiplying $H_k$ to the aggregated signal $S^k \rmX(t-k)$ can be computed locally for any matrix $H_k$.

\subsubsection{Graph Neural Networks}
A graph convolution as defined above can be viewed as a two-step process: first, a set of delayed information is aggregated at each node using the graph shift operator $S$; then, the aggregated information is used to synthesize an output by using the filter weights $\{H_k\}$. GNNs \citep{bruna2013spectral, kipf2016semi, atwood2015diffusion, gama2018convolutional} seek to expand the expressiveness of graph convolutions by repeatedly applying one or both steps in combination with nonlinearities.

One architecture that achieves this is delayed aggregation GNNs (AGNNs) \citep{gama2018convolutional, tolstaya2020learning}.  Informally, an AGNN first aggregates information through a graph convolution over the states. The output is then fed through a standard convolution neural network (CNN).  As the CNN is acting on signals synthesized from locally available information, such an architecture naturally encodes the information sharing specified by the graph shift operator of the graph convolution layer.  In the interest of space, we omit a detailed description of AGNNs, and instead refer the reader to \citet{gama2020graph} for a more detailed overview of their use in the context of distributed control.

When information exchange constraints are defined only in terms of spatial constraints, and not delays, one can repeated apply both the aggregation and the convolution steps to parameterize the corresponding distributed controller using a graph convolution neural network (GCNN). A GCNN controller with $L$ layers can be written recursively as
\begin{equation}
\begin{aligned}
    \rmU(t) &= X_{L} \\
    X_{l} &= \sigma \left( \sum_{k=0}^{K_l} S^k X_{l-1} H_{l,k} \right),\quad l=1,...,L \\
    X_{0} &= \rmX(t),
\end{aligned}
\label{eq: GNN}
\end{equation}
where $\sigma$ is applied element-wise and $H_{l,K}$ is a set of trainable parameters. Note that while the application of the graph shift operator maintains locality of information exchange, the sub-controller at subsystem $i$ will need to instantaneously collect the local information $\infoset_{j,t}$ from all neighboring subystems $j$ satisfying $\dist^{\mathcal{G}_c}(i,j)\leq\sum_{l=0}^{L} K_{l}$ -- thus one must be careful to balance the depth $L$ of the network, the density of the graph shift operator $S$, and the aggregation horizon $K_l$, to ensure that locality is indeed preserved.  The size of parameter matrices $H_{l,k} \in \R^{r_l \times c_l}$ can be arbitrary as long as internal state dimensions are consistent, i.e., $r_l = c_{l-1}$ and $r_{0} = p, c_{L} = q$. The total number of parameters of a GNN is thus a tunable hyperparameter totaling $\sum_{l=0}^{L} (K_l + 1) r_{l} c_{l}$.  We end by noting that distributed controllers parameterized as GCNNs will outperform comparable\footnote{By comparable, we mean with comparably chosen hyper-parameters such that the overall number of free parameters of the function classes are approximately the same.} delayed AGNNs defined in terms of the same graph shift operator $S$, because the former does not enforce delay constraints on information sharing, whereas the latter does.  As such, we use GCNN-based controllers as a baseline for comparison in our experiments.

\subsubsection{Graph Recurrent Neural Networks} \label{sec: GRNN}
For the problem of communication topology co-design, we propose parameterizing distributed controllers using GRNNs -- GRNNs have been successfully used for decentralized control in the context of imitation learning in \cite{gama2020graph}. GRNNs extend the aforementioned GNN-based controller parameterizations by introducing a local hidden state $z_i(t) \in \R^{h}$ at each sub-controller, resulting in a \emph{dynamic} distributed controller.  We further allow sub-controllers to communicate both their local state and internal state with neighboring subsystems, i.e., in this parameterization, the local information set of subsystem $i$ is given by $\infoset_{i,t}:=\{x_i(t),z_i(t)\}$. Let  $\rmZ(t):=(z_i(t)^\top)_{i=1}^N \in \R^{N \times h}$ denote the full internal state obtained by stacking the subsystem internal states as in equation (\ref{eq: stacked-states}).  
While more general forms of GRNNs exist and can be used, we propose using the following compact GRNN parameterization, where an update step can be computed as

\begin{equation}
\begin{aligned}
    \rmZ(t) &= \sigma(S\;\rmZ(t-1)\;W + \rmX(t)\; F), \\
    \rmU(t) &= \rmZ(t)\;G,
\end{aligned}
\label{eq: RGNN}
\end{equation}
where $F \in \R^{p \times h}, G \in \R^{h \times q}, W \in \R^{h \times h}$ are trainable parameters.  

We selected this architecture because it enjoys certain properties that make it naturally suited for encoding and optimizing communication delay structure.  At each time step, the update step \eqref{eq: RGNN} applies the graph shift operator $S$ once on the internal states, which ensures that information within the controller is consistent with information flow across the communication network.  Thus a controller parameterized by equation \eqref{eq: RGNN} satisfies the information sharing constraints defined in \eqref{eq:total_info} and (\ref{eq: info constraint}) with the information set $\infoset_{i,t} = \{x_i(t), z_i(t)\}$.\footnote{We note that $x_i(t)$ is in fact only ever needed at subsystem $i$, but we include it in $\infoset_{i,t}$ to be consistent with equations \eqref{eq:total_info} and \eqref{eq: info constraint}.}  Further, the parameterization \eqref{eq: RGNN} is an affine mapping of the graph shift operator $S$ composed with a point-wise nonlinearity $\sigma$.  The simplicity with which $S$ appears in parameterization \eqref{eq: RGNN} makes it well suited for treating $S$ as an optimization variable.  In particular, whereas existing GNN-based distributed controllers \citep{tolstaya2020learning,gama2020graph,khan2020graph,gama2021distributed} assume a \emph{fixed} graph shift operator $S$, we take a slight departure and \emph{jointly optimize} the GRNN-based distributed controller parameter values and graph shift operator sparsity pattern and parameter values.  Doing so allows us to co-design a distributed controller and the communication graph topology $\mathcal{G}_c$ needed to implement it.

%% file: tex/3-Formulation.tex
\section{Co-Design via Regularized Empirical Risk Minimization}
Given a distributed linear system (\ref{eq: subsystem linear dynamics}), our goal is to explore the design space of distributed controllers satisfying communication constraints of the form (\ref{eq: info constraint}) while minimizing a cost function over a finite horizon  $T \in \mathbb{Z}^{+}$. We consider two problem formulations. First, we show how to solve the distributed controller design problem on a \emph{given} communication network $\mathcal{G}_c$ using GRNN with an empirical risk minimization (ERM) approach akin to that proposed in \citet{gama2021distributed}. Next, we build on this formulation and show how to jointly design a distributed controller and the communication topology $\mathcal{G}_c$ needed to implement it through the use of $\ell_1$-regularized ERM. 


\subsection{Distributed GRNN Control Design: Given Topology}
We consider the linear system (\ref{eq: joint linear dynamics}), which can be equivalently rewritten as
\begin{equation}
    \vectorize{\rmX(t+1)} = A\;\vectorize{\rmX(t)} + B\;\vectorize{\rmU(t)}.
    \label{eq: stacked linear dynamics}
\end{equation}
Let $\mathcal X \subseteq \R^{N\times p}$ be a set, $\mathcal D$ be a distribution over $\mathcal X$, and the initial conditions of our system be distributed as $\rmX(0)\sim{}\mathcal{D}$.  Suppose the quality of a system trajectory is characterized by the cost function $J\left (\{\rmX(t)\}_{t=0}^{T}, \{\rmU(t)\}_{t=0}^{T} \right)$. 
We can then pose our optimal control problem as
\begin{equation}
    \begin{array}{rl}
    \minimize_{\{\gamma_{i,t}\}} 
    & \mathbb{E}_{\rmX(0) \sim \mathcal D} \left[ J\left (\{\rmX(t)\}_{t=0}^{T}, \{\rmU(t)\}_{t=0}^{T} \right) \right]\\
    \textrm{s.t.} & \text{dynamics \eqref{eq: stacked linear dynamics}}, \\
    & u_i(t) = \gamma_{i,t}(\totalinfoset_{i,t}(\mathcal{G}_c)),\; i=1,...,N,\, t=0,...,T
\end{array}
\label{eq: P1 problem}
\end{equation}
In general, this problem is non-convex even if the cost function is quadratic and the distribution $\mathcal D$ is Gaussian (corresponding to LQG control) due to the communication constraints \eqref{eq:total_info} and (\ref{eq: info constraint}) (encoded as $u_i(t) = \gamma_{i,t}(\totalinfoset_{i,t}(\mathcal{G}_c))$). We thus aim to approximate the optimal solution to problem \eqref{eq: P1 problem} by using the GRNN parameterization \eqref{eq: RGNN} for distributed controllers, and leveraging ERM. 

As presented in Section \ref{sec: GRNN}, GRNN models can encode the communication constraints \eqref{eq:total_info} and (\ref{eq: info constraint}) induced by topology $\mathcal{G}_c$ by restricting the graph shift operator to satisfy $S \in \mathcal{S}_{\mathcal{G}_c},$ i.e., by enforcing that $S$ be a valid graph shift operator for the communication topology $\mathcal{G}_c$.  In practice, we enforce this constraint by restricting the support of $S$ to be consistent with that of the adjacency matrix of the graph $\mathcal{G}_c$. Thus, we draw initial conditions  $\{\rmX_{p}(0)\}_{p=1}^{k} \sim{} \mathcal D^k$ and pose the distributed GRNN control design problem as the following ERM problem:
    

\begin{equation}
    \begin{array}{rl}
     \underset{S, F, G, W}{\minimize} &
     \frac{1}{k} \sum_{p=0}^{k} \left[ J \left( \{\rmX_{p}(t)\}_{t=0}^{T}, \{\rmU_{p}(t)\}_{t=0}^{T} \right) \right]\\
    \textrm{s.t.} & \text{dynamics (\ref{eq: stacked linear dynamics}),} \\
    & \rmZ_p(t) = \sigma(S\;\rmZ_p(t-1)\;W + \rmX_p(t)\; F), \\
    &\rmU_p(t) = \rmZ_p(t)\;G, \, p=0,...,k, \\
    &S \in \mathcal{S}_{\mathcal{G}_c}, \, (S,F,G,W) \in \Theta,
    \end{array}
    \label{eq: P1 ERM}
\end{equation}
where the set $\Theta$ is a user-specified set constraining the overall expressivity of the model to prevent over-fitting.  A typical example would to impose Frobenius norm bounds on the parameters, i.e., $\Theta = \{(S,F,G,W) \, : \, \|S\|_F^2 \leq B_S, \|F\|_F^2 \leq B_F,  \|G\|_F^2 \leq B_G,\|W\|_F^2 \leq B_W\}.$

Following \citet{gama2021distributed}, this problem can be approximately solved in a self-supervised manner using stochastic gradient descent. During training, we simulate the initial conditions forward with the controllers parameterized with the current parameter iterates $\{\hat S, \hat F, \hat G, \hat W\}$, and compute their cost. The gradient of the cost with respect to the parameters can then be found efficiently via back propagation through time \citep[\S10.2.2]{goodfellow2016deep}.  We note that in contrast to prior work \citep{tolstaya2020learning,gama2020graph,khan2020graph,gama2021distributed}, we propose optimizing the parameters of the graph shift operator $S$ in addition to the parameters $\{W,F,G\}$.

\subsection{Communication Topology Co-Design}
Unlike the previous setting, here we assume that the topology $\mathcal{G}_c$ is also to be designed.  In particular, we aim to develop a methodology that allows us to efficiently explore the tradeoff space between communication complexity, as measured by the cardinality of the edge set $|\mathcal{E}_c|$, and controller performance. To do so, we propose a two step process.


First, we co-design the communication topology $\mathcal{G}_c$ by solving the $\ell_1$-regularized ERM problem
\begin{equation}
    \begin{array}{rl}
    \underset{S, F, G, W}{\minimize} &
     \frac{1}{k}  \sum\limits_{p=0}^{k} \left[ J \left( \{\rmX_{p}(t)\}_{t=0}^{T}, \{\rmU_{p}(t)\}_{t=0}^{T} \right) \right] + \lambda \|S\|_1 \\
    \textrm{s.t.} & \text{dynamics (\ref{eq: stacked linear dynamics}),} \\
    & \rmZ_p(t) = \sigma(S\;\rmZ_p(t-1)\;W + \rmX_p(t)\; F), \\
    &\rmU_p(t) = \rmZ_p(t)\;G, \, p=0,...,k, \\
    &(S,F,G,W) \in \Theta.
    \end{array}
    \label{eq: P2 step 1}
\end{equation}
Note that in contrast to optimization problem \eqref{eq: P1 ERM}, we have dropped the constraint that the graph shift operator satisfy $S \in \mathcal{S}_{\mathcal{G}_c}$, and instead added an $\ell_1$ regularizer to the objective function, which is known to promote sparsity \citep{tibshirani1996regression,donoho2006compressed}.   
Thus, by varying the regularization weight $\lambda$, we can trade off sparsity of the network against controller performance. 

Denote by $\hat{S}(\lambda)$ the solution to optimization problem \eqref{eq: P2 step 1} for a given value of $\lambda$, and let $\hat{A}(\lambda) = \mathrm{supp}(\hat{S}(\lambda))$ be the corresponding adjacency matrix defined by its support.\footnote{Note that in practice, we apply a thresholding procedure to determine adjacency matrix, i.e., $\hat{A}_{ij} = 1$ if an only if $|\hat S_{ij}|\geq \varepsilon$, for some small numerical threshold $\varepsilon$.}  The designed graph $\hat{\mathcal G}(\lambda)$ is then taken to be the graph induced by the adjacency matrix $\hat{A}(\lambda)$.  We then proceed with a standard \emph{refinement step}, and solve optimization problem \eqref{eq: P1 ERM} subject to information sharing constraints imposed by the designed communication topology $\hat{\mathcal{G}}(\lambda)$. 

%% file: tex/4-Experiments.tex
\section{Numerical Experiments}
\label{sec: experiments}
All code needed to reproduce the examples found in this section will be made at available at the following repository: \href{https://github.com/unstable-zeros/grnn-comms-codesign}{https://github.com/unstable-zeros/grnn-comms-codesign}.

We consider the distributed linear quadratic regulator (LQR) problem over $N=20$ subsystems and a finite time horizon $T=50$. Specifically, we consider the case where the state and control actions of the subsystems are scalars, i.e., $p=q=1$. Under this assumption, $\rmX(t)=x(t), \rmU(t) = u(t)$, and the problem can be written as
\begin{equation}
    \begin{array}{rl}
    \minimize  &  \mathbb E_{x(0)\sim{}\N(0,I)}\left[\sum_{t=0}^{T-1} x(t)^{\intercal} Q x(t) + u(t)^{\intercal} R u(t) + x^{\intercal}(T) P x(T) \right] ,\\
    \textrm{s.t.}  & x(t+1) = Ax(t) + Bu(t),\; t=0,1,...,T-1, \\
     & u_i(t) = \gamma_{i,t}(\totalinfoset_{i,t}(\mathcal{G}_c)) \\ & \quad \quad \text{for $i=1,\dots,N,\, t=0,\dots,T$},
    \end{array}
    \label{eq: LQR}
\end{equation}
for dynamics matrices $A, B \in \R^{N \times N}$, $N \times N$ symmetric cost matrices $Q\succeq 0, R\succ 0, P\succeq 0$, and communication topology $\mathcal{G}_c$. 

\textbf{Generating Problem Instances:}  An instance of the LQR problem (\ref{eq: LQR}) is fully defined by the tuple $(\mathcal{G}_c,A, B, Q, R, P)$. We generate problem instances using the same process for all experiments. For each instance, we start by creating a communication topology $\mathcal{G}_c$ by randomly sampling $N$ numbers $\{u_i\}_{i=1}^N\sim{}U[0,1]$, and creating a bi-directional link between $v_i$ and each of its $5$ nearest points as defined by the topology on the interval $[0,1]$ under the metric $d(v_i,v_j)=|u_i-u_j|$.  Following \citet{gama2021distributed}, we generate the dynamics matrices $A$ and $B$ to share the same eigenvectors as the normalized adjacency matrix of $\mathcal{G}_c$, and sample their eigenvalues i.i.d.\ from the standard normal distribution -- hence both $A$ and $B$ are symmetric matrices. Then, to induce further sparsity in the dynamics, we set entries $A_{ij} = 0$ and $B_{ij} = 0$ whenever $\dist^{\mathcal{G}_c}(i \rightarrow j) > 3$. Finally, we normalize the matrices $A$ and $B$ to have the prescribed norms for the experiments being run. We take the cost matrices $Q = R = I_{N}$. Finally, the terminal state cost matrix $P$ is taken to be the solution to the Discrete Algebraic Riccati Equation (DARE)
\begin{equation}
    A^\intercal P A - P - A^\intercal P B (B^\intercal P B + R)^{-1}B^\intercal P A = -Q.
    \label{eq: dare}
\end{equation}
This ensures that the centralized solution is stabilizing.  We end by noting that communication topologies $\mathcal{G}_c$ generated this way have, on average, $141$ directed edges (out of a possible 400).

In all of the following experiments, we use the ADAM algorithm \citep{kingma2014adam} to approximately solve the resulting ERM problems with a batch size of $20$, and forgetting factors of $0.9$ and $0.999$. For the GRNN models, we choose the learning rate to be $0.02$, and apply cosine annealing on top of ADAM for learning rate scheduling. For the GCNN models, we use a learning rate of $0.01$, and decay the learning rate by a factor of $0.9$ every $10$ batches, as specified in \citet{gama2020graph}. The initial conditions in each batch are sampled i.i.d.\ from $\N(0,I_N)$. We train the models with $750$ batches of initial conditions. 

\subsection{Performance Benchmarks}
In this set of experiments, we compare the performance of different variants of the GRNN parameterization \eqref{eq: RGNN} with a GCNN benchmark from \citet{gama2021distributed} on the LQR problem.  We note that we focus exclusively on GNN based controllers in our study, as \citet{gama2021distributed} has already conducted extensive studies comparing GNN-based distributed controllers against more traditional (fully-connected) neural network parameterizations of distributed controllers.

\textbf{Architectures:} We consider four different variants of the GRNN-based controller \eqref{eq: RGNN}. All the GRNN variants considered share the same architecture with a hidden state dimension of $5$. However, they differ in their training objective functions and constraints.
\begin{itemize}
    \item \textit{(GRNN)} approximately solves the LQR problem \eqref{eq: LQR} using the ERM problem (\ref{eq: P1 ERM}). 
    \item \textit{(GRNN-Dense)} does not place constraints on the graph shift operator $S$, and can thus design a centralized controller. 
    \item \textit{(GRNN-Sparse)} co-designs a communication topology and distributed controller using (\ref{eq: P2 step 1}) to identify a topology $\hat{\mathcal{G}}$ and (\ref{eq: P1 ERM}) restricted to the identified topology for the subsequent refinement step\footnote{We take the threshold value for determining the adjacency matrix support to be $\varepsilon = 0.004$.}. 
    \item \textit{(GRNN-Fixed)} restricts $S$ to be constant and only optimizes over the parameters $F, G$, and $W$. 
    \item \textit{GCNN} is the GCNN architecture in \citet{gama2021distributed}, which we take as a benchmark.  We take the graph shift operator $S$ to be the normalized adjacency matrix of $\mathcal{G}_c$. The GCNN model has $2$ layers with $5$ and $1$ filter taps, respectively and uses a total of $192$ parameters.
\end{itemize}

\textbf{Training and evaluation:} We run our experiments across $50$ random LQR instances.  We report the costs achieved by each controller normalized with respect to that achieved by the centralized LQR optimal controller $K_{LQR} = - (B^\intercal P B)^{-1}B^\intercal P A$ computed via the solution to the DARE \eqref{eq: dare}.  We note that the centralized cost represents a fundamental lower limit on the achievable performance, and thus a normalized cost of $1$ is optimal.

\begin{figure*}[t]
     \centering
     \includegraphics[width=0.45\textwidth]{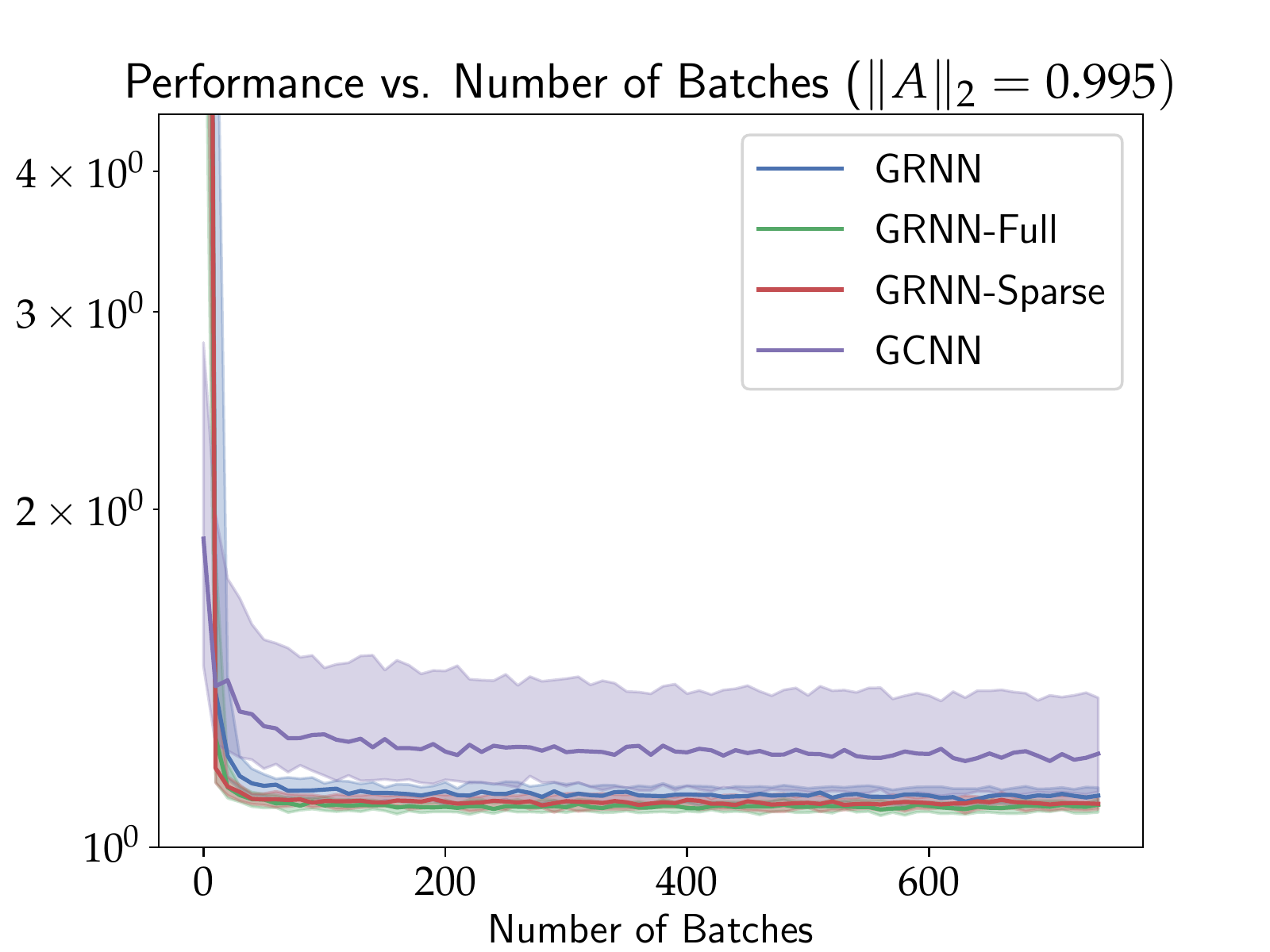}
    ~~~~~~\includegraphics[width=0.45\textwidth]{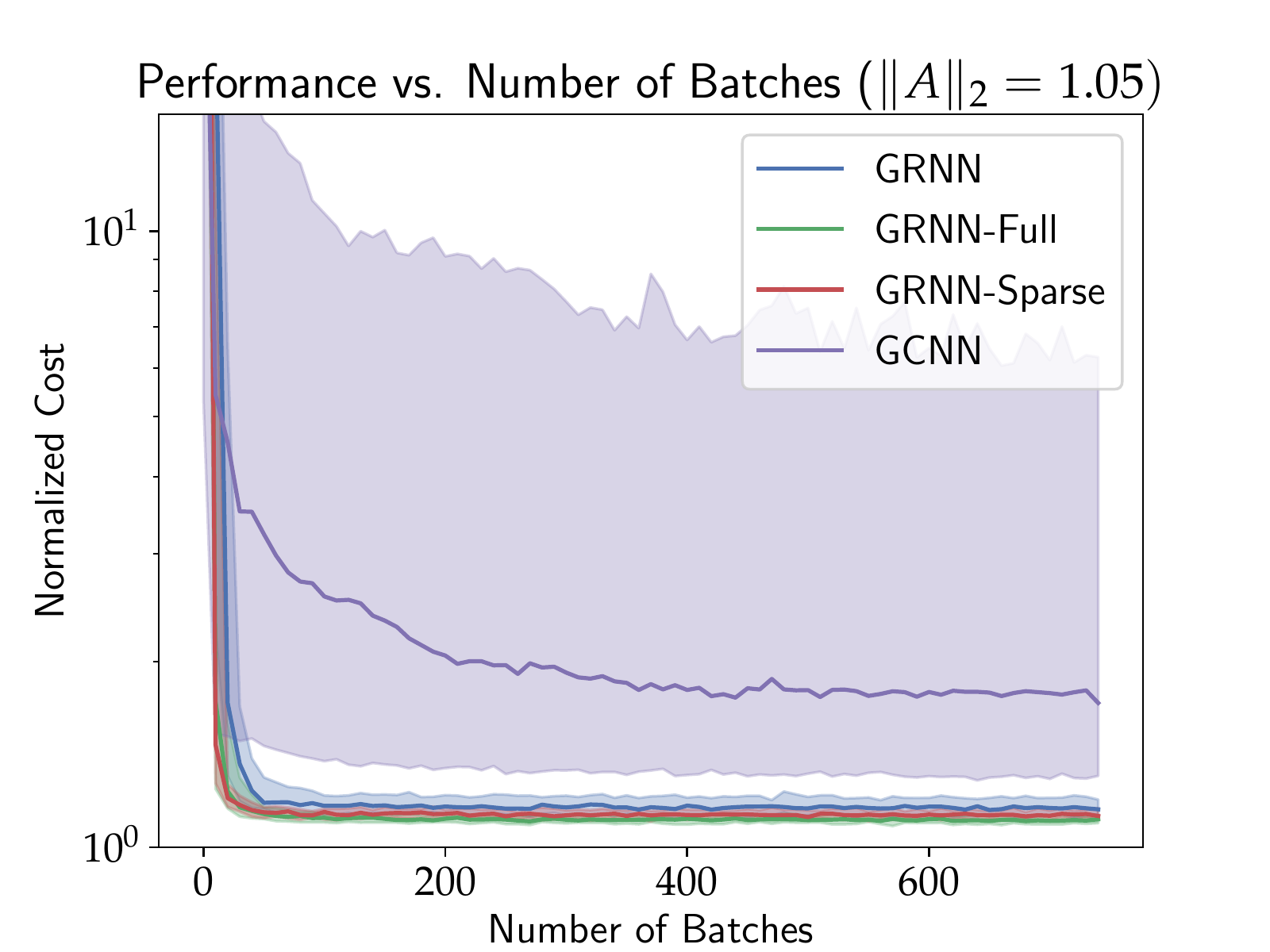}
    \caption{Training curve for the respective distributed controller parameterizations for problem instances with $\|A\|_2=0.995$ (Left) and $\|A\|_2=1.05$ (Right).  The solid line denotes the median of normalized costs, and the shaded regions denotes quartiles, over $50$ randomly generated problem instances.
    }
             \label{fig:benchmark}
\end{figure*}

 We run this experiment on $50$ randomly generated pairs of communication networks $\mathcal{G}_c$ and dynamics matrices $(A,B)$. Every $10$ batches, we evaluate the controllers over a validation set of $100$ initial conditions and report the average performance.  To test the performance of the GRNN method on both stable and unstable environments, we repeat the process on two scenarios, one with the norm of the dynamics matrix $\|A\|_2=0.995$, the other with $\|A\|_2=1.05$. When training the GRNNs, we also apply weigh-decay (WD) (Frobenius-norm regularization) on the parameters $(S, F, W, G)$, when applicable.  When training the GCNNs, we experimented with no WD on the filter-taps, as specified in \citet{gama2021distributed}, as well as with weight-decay: we found the results with and without WD to be qualitatively and quantitatively similar, and present here only those without WD.  When solving the co-design problem \eqref{eq: P2 step 1}, we set the regularization penalty $\lambda$ to $1$ and $2$ for $\|A\|_2=0.995$ and $\|A\|_2=1.05$, respectively: these values were selected based off of the co-design experiments we describe next. We report the median relative cost for all methods in Table \ref{tab:benchmark}, and plot the learning curves of the controllers during the training in Figure \ref{fig:benchmark}.

\textbf{Discussion:} First, we note that the GRNN parametrization outperforms the the GCNN benchmark on both the stable and unstable dynamics despite the latter parameterization \emph{not enforcing communication delay constraints}. This suggests that the GRNN model can leverage the recurrent structure to aggregate information without needing a large number of filter taps.  We further observe that all GRNN architectures enjoy reduced variance as compared to the GCNN architecture\footnote{The GCNN architecture with WD has only slightly less variance at the expense of even worse performance.}. Interestingly, we also observe that the GRNN-Sparse controllers, for which the communication topology is co-designed, are able to outperform the GRNN models with a prescribed communication topology that is designed to be well aligned with the system dynamics (via their eigenspaces). Furthermore, the GRNN-Sparse controllers use on average, $53.4$ and $56.1$ directed communication links for the stable and unstable dynamics, respectively, far fewer than the average number of links in the given communication topology. This indicates that our co-design algorithm was able to identify key links in the communication topology needed for control, and further suggests that by restricting the controller parameterization to a simpler (sparser) model class, less data is required to learn a high-performing controller.  We leave formalizing these observations in the context of the ERM framework to future work.


\begin{table}
    \centering
    \begin{tabular}{ |c|c|c| } 
    \hline
    Architecture & $\|A\|_2=.995$ & $\|A\|_2=1.05$ \\ 
    \hline \hline
    Autonomous  &2.534 &69.294 \\ 
    \hline
    GRNN      &1.112 & 1.153\\ 
    \hline
    GRNN-Dense  &1.091 &1.110\\ 
    \hline
    GRNN-Sparse &1.093 &1.124\\
    \hline
    GRNN-Fixed  &1.267 &2.210\\ 
    \hline
    GCNN   &1.212 &1.715\\ 
    \hline
    \end{tabular}
    \caption{Median normalized cost for the perfomrance benchmarking experiments described in \S\ref{sec: experiments}-A over $50$ randomly generated problem instances.  The GCNN baseline is that described in \citet{gama2021distributed}.}
    \label{tab:benchmark}
\end{table}

\subsection{Controller-Topology Co-design}
\begin{figure}[h]
     \centering
     \includegraphics[width=0.5\textwidth]{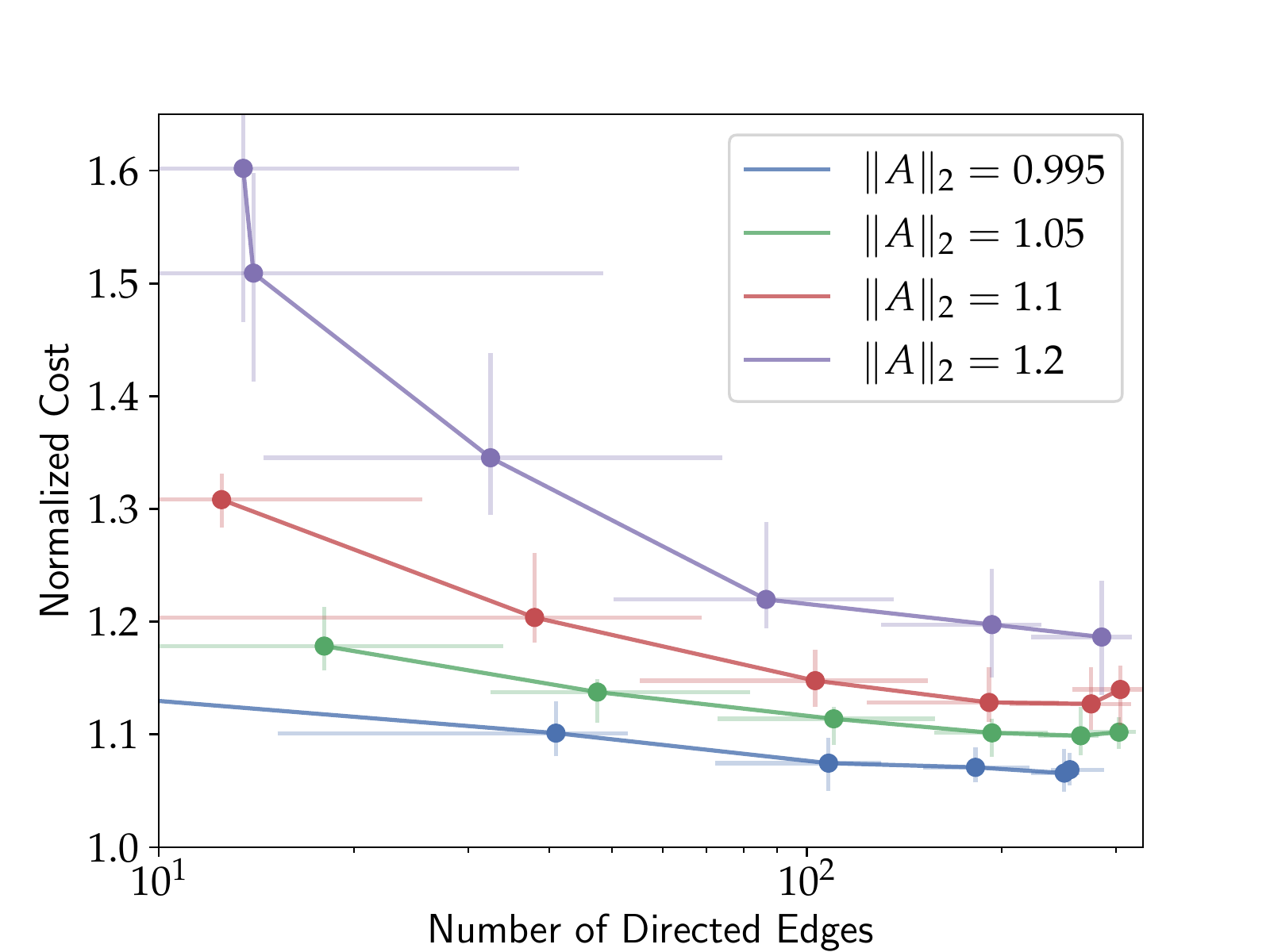}
     \caption{Trade-off curves of the normalized cost versus communication topology complexity (characterized via the number of directed edges $|\hat{\mathcal E}|$) for varying values of $\|A\|_2$.  For each value of $\|A\|_2$, we plot the median values across $30$ randomly generate problem instances, with horizontal and vertical error bars denoting the quartiles in number of edges and normalized cost, respectively.}
     \label{fig:tradoff}
 \end{figure}

\begin{figure*}
    \centering
    ~~~\begin{tabular}{@{}c@{}}
        \centering
        \subfloat[Physical interaction topology $\mathcal{G}_{dyn}$]{
            \includegraphics[width=0.2\textwidth]{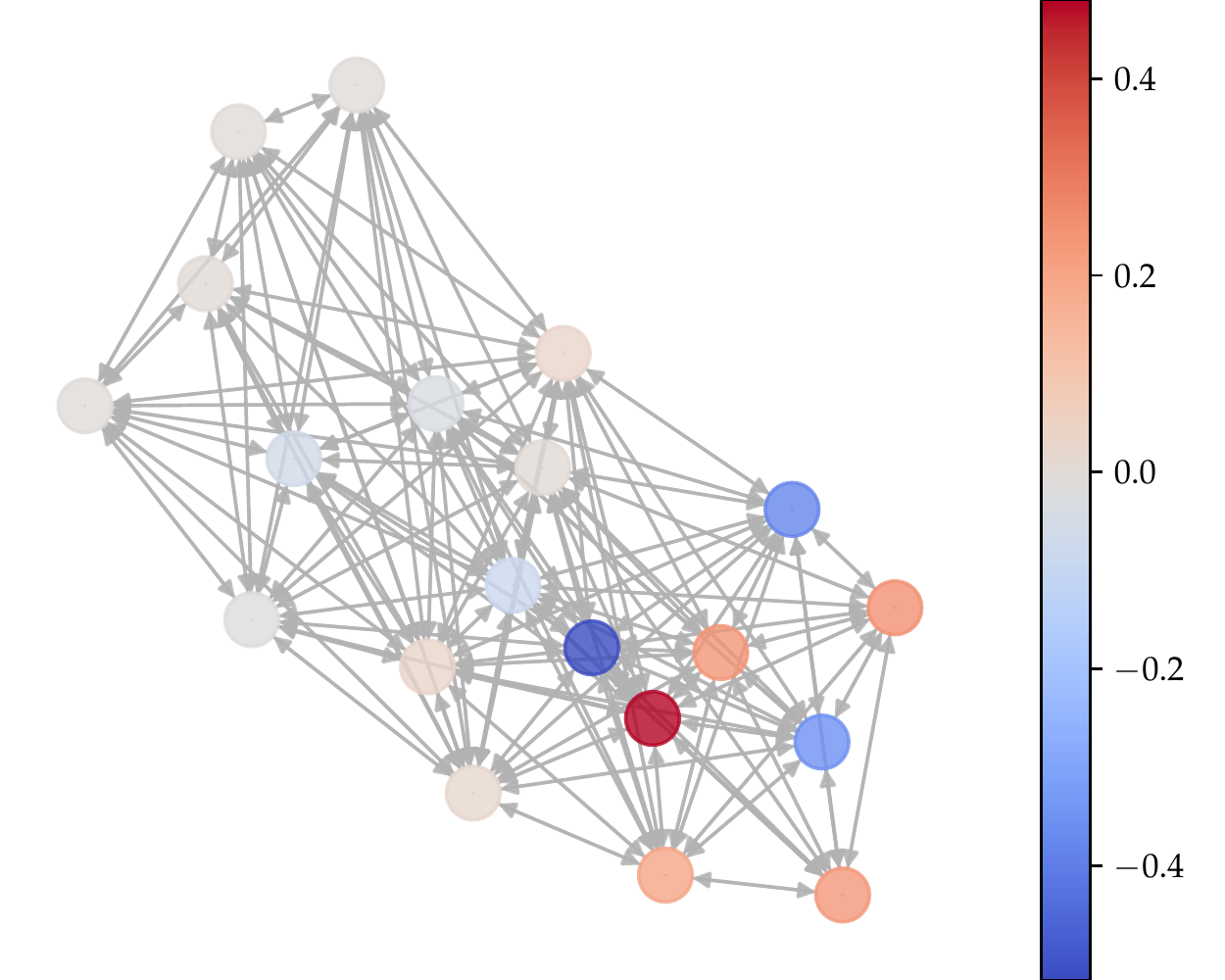}
        }
    \end{tabular}~~~
    \begin{tabular}{@{}c@{}}
        \subfloat[Designed topologies $\hat{\mathcal{G}}(\lambda)$]{
            \includegraphics[width=0.7\textwidth]{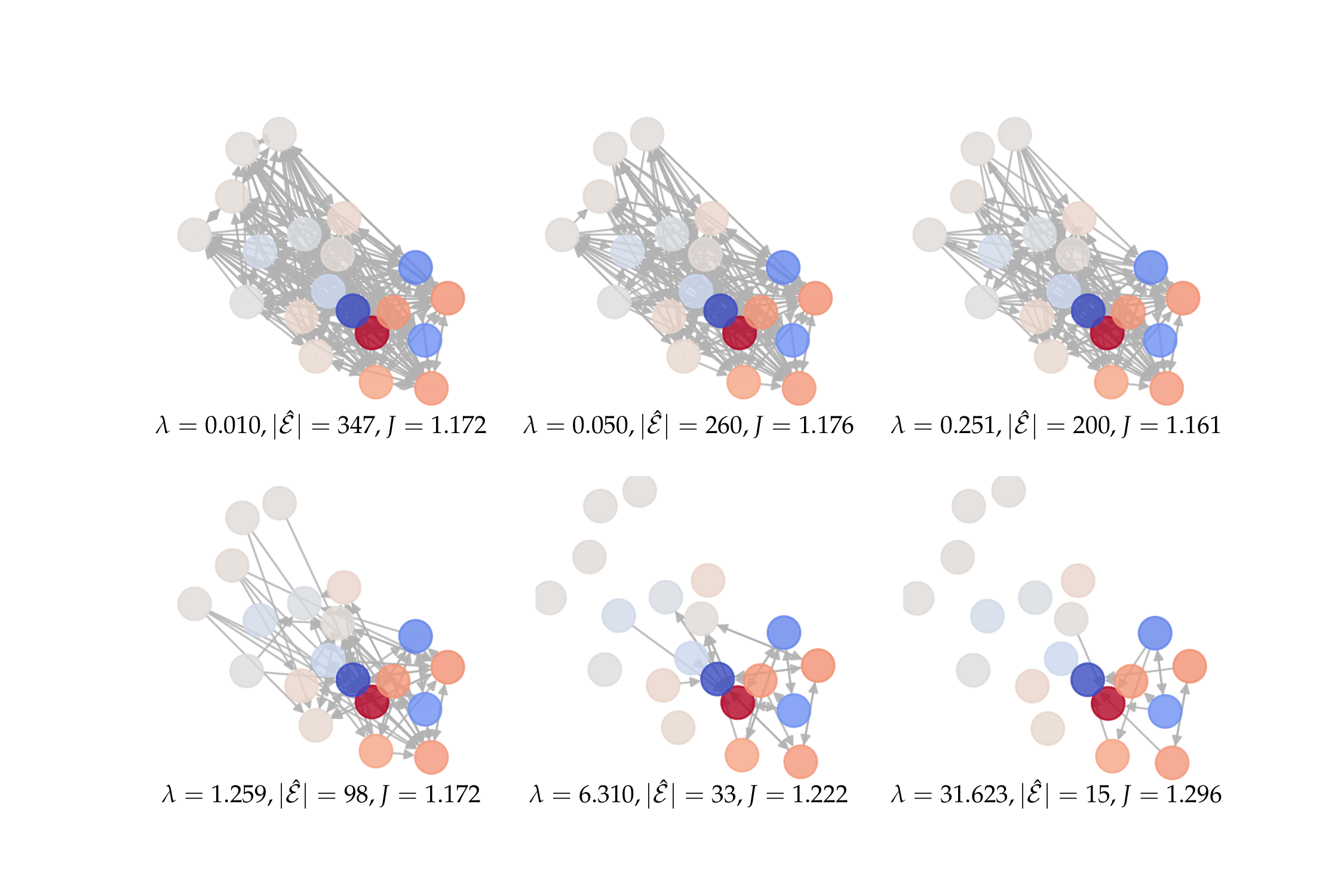}
        }
    \end{tabular}~~~
    \caption{A visualization of the evolution of the designed topology $\hat{\mathcal{G}}(\lambda)$ as the regularization parameter $\lambda$ is varied.  Under each identified topology, we indicate the regularization parameter value $\lambda$, the identified edge set cardinality $|\hat{\mathcal{E}}|$, and the normalized cost achieved by the co-designed controller $J$.  For this example, $\|A\|_2 = 1.1$. The color of the nodes represent the eigenvector of $A$ associated with the leading eigenvalue $\lambda = -1.1$, with nodes with large values  colored more prominently.  As can be seen, as the graph becomes more sparse, links between these nodes remain.}
    \label{fig:graphs}
\end{figure*}

We show how our approach allows us to efficiently explore the trade-off between the sparsity of the communication topology of a distributed controller and the performance that it achieves via our co-design problem \eqref{eq: P2 step 1}. We run our GRNN co-design algorithm for four families of LQR instances, each characterized by the operator norm of the $A$ matrix defining the system dynamics, with $\|A\|_2$ ranging from stable (less than 1) to unstable (greater than 1). For each value of $\|A\|_2$, we solve the co-design problem \eqref{eq: P2 step 1} with regularization weight $\lambda$ varying across several orders of magnitude, sweeping out a curve $\hat{\mathcal{G}}(\lambda)$ in communication topology space: a representative example of such a curve is illustrated in Fig.~\ref{fig:graphs}. We record the cost achieved by the co-designed distributed controller, as well as number of directed links needed to implemented it as captured by the cardinality of the designed edge set $|\mathcal{E}_c|$ of the communication topology over $30$ realizations for each value of $\|A\|_2$. We plot the resulting trade-off curves in Figure \ref{fig:tradoff}.

\textbf{Discussion:} As the results show, there is a clear trade-off between the sparsity of the communication topology and the performance of the system. We observe that the trade-off becomes more pronounced as the norm of $A$ grows, i.e., as the open-loop system becomes more unstable and consequently, difficult to control.  Interestingly, we observe that for all values of $\|A\|_2$, our co-design algorithm is able to explore the entire design space from dense (virtually centralized) to (nearly) completely decentralized.  We emphasize that such (nearly) completely decentralized communication topologies do not satisfy the conditions needed for convexity/tractability of linear distributed optimal control such as quadratic invariance \citep{rotkowitz2005characterization,rotkowitz2010convexity} or funnel causality \citep{bamieh2005convex}, and yet reasonably well-performing distributed optimal controllers are identified.  While prior work \citep{matni2016regularization} identifies conditions under which such co-designed architectures are optimal, we are not aware of such conditions in the context of data-driven distributed control.  We view our results as strong empirical evidence that such an investigation should be undertaken.



%% file: tex/5-conclusion.tex
\section{Conclusion \& Future Work} \label{sec: conclusion}
Jointly designing communication topologies with distributed controllers is difficult in general. In this work, we proposed a data-driven method to solve the co-design problem by parametrizing the distributed controllers as a GRNN and solving an $\ell_1$-regularized ERM problem that allows for a principled tradeoff between control cost and communication topology complexity. We demonstrated empirically that our proposed GRNN parametrization can generate distributed controllers with good performance and our method allows us to efficiently explore the trade-off space between the sparsity of the designed communication network and the performance of the distributed controllers.  Future work will look to provide generalization bounds on the learned controllers by leveraging recent results that allow for uniform convergence guarantees to be applied to dynamical systems under suitable stability assumptions \citep{tu2021closing} along with quadratic constraint based conditions for the stability of dynamical systems modelled as recurrent neural networks \citep{revay2020convex}.

%% file: main.bbl
\begin{thebibliography}{30}
\providecommand{\natexlab}[1]{#1}
\providecommand{\url}[1]{\texttt{#1}}
\expandafter\ifx\csname urlstyle\endcsname\relax
  \providecommand{\doi}[1]{doi: #1}\else
  \providecommand{\doi}{doi: \begingroup \urlstyle{rm}\Url}\fi

\bibitem[Pequito et~al.(2015)Pequito, Kar, and Pappas]{pequito2015minimum}
S{\'e}rgio Pequito, Soummya Kar, and George~J Pappas.
\newblock Minimum cost constrained input-output and control configuration
  co-design problem: A structural systems approach.
\newblock In \emph{2015 American control conference (ACC)}, pages 4099--4105.
  IEEE, 2015.

\bibitem[Tzoumas et~al.(2016)Tzoumas, Jadbabaie, and Pappas]{tzoumas2016sensor}
Vasileios Tzoumas, Ali Jadbabaie, and George~J Pappas.
\newblock Sensor placement for optimal kalman filtering: Fundamental limits,
  submodularity, and algorithms.
\newblock In \emph{2016 American Control Conference (ACC)}, pages 191--196.
  IEEE, 2016.

\bibitem[Summers and Lygeros(2014)]{summers2014optimal}
Tyler~H Summers and John Lygeros.
\newblock Optimal sensor and actuator placement in complex dynamical networks.
\newblock \emph{IFAC Proceedings Volumes}, 47\penalty0 (3):\penalty0
  3784--3789, 2014.

\bibitem[Summers et~al.(2015)Summers, Cortesi, and
  Lygeros]{summers2015submodularity}
Tyler~H Summers, Fabrizio~L Cortesi, and John Lygeros.
\newblock On submodularity and controllability in complex dynamical networks.
\newblock \emph{IEEE Transactions on Control of Network Systems}, 3\penalty0
  (1):\penalty0 91--101, 2015.

\bibitem[Lin et~al.(2013)Lin, Fardad, and Jovanovi{\'c}]{lin2013design}
Fu~Lin, Makan Fardad, and Mihailo~R Jovanovi{\'c}.
\newblock Design of optimal sparse feedback gains via the alternating direction
  method of multipliers.
\newblock \emph{IEEE Transactions on Automatic Control}, 58\penalty0
  (9):\penalty0 2426--2431, 2013.

\bibitem[Fardad et~al.(2014)Fardad, Lin, and Jovanovi{\'c}]{fardad2014design}
Makan Fardad, Fu~Lin, and Mihailo~R Jovanovi{\'c}.
\newblock Design of optimal sparse interconnection graphs for synchronization
  of oscillator networks.
\newblock \emph{IEEE Transactions on Automatic Control}, 59\penalty0
  (9):\penalty0 2457--2462, 2014.

\bibitem[Matni and Chandrasekaran(2016)]{matni2016regularization}
Nikolai Matni and Venkat Chandrasekaran.
\newblock Regularization for design.
\newblock \emph{IEEE Transactions on Automatic Control}, 61\penalty0
  (12):\penalty0 3991--4006, 2016.

\bibitem[Dhingra et~al.(2014)Dhingra, Jovanovi{\'c}, and Luo]{dhingra2014admm}
Neil~K Dhingra, Mihailo~R Jovanovi{\'c}, and Zhi-Quan Luo.
\newblock An admm algorithm for optimal sensor and actuator selection.
\newblock In \emph{53rd IEEE Conference on Decision and Control}, pages
  4039--4044. IEEE, 2014.

\bibitem[Chandrasekaran et~al.(2012)Chandrasekaran, Recht, Parrilo, and
  Willsky]{chandrasekaran2012convex}
Venkat Chandrasekaran, Benjamin Recht, Pablo~A Parrilo, and Alan~S Willsky.
\newblock The convex geometry of linear inverse problems.
\newblock \emph{Foundations of Computational mathematics}, 12\penalty0
  (6):\penalty0 805--849, 2012.

\bibitem[Matni(2015)]{matni2015communication}
Nikolai Matni.
\newblock Communication delay co-design in $h_2$-distributed control using
  atomic norm minimization.
\newblock \emph{IEEE Transactions on Control of Network Systems}, 4\penalty0
  (2):\penalty0 267--278, 2015.

\bibitem[Rotkowitz and Lall(2005)]{rotkowitz2005characterization}
Michael Rotkowitz and Sanjay Lall.
\newblock A characterization of convex problems in decentralized control.
\newblock \emph{IEEE transactions on Automatic Control}, 50\penalty0
  (12):\penalty0 1984--1996, 2005.

\bibitem[Rotkowitz et~al.(2010)Rotkowitz, Cogill, and
  Lall]{rotkowitz2010convexity}
Michael Rotkowitz, Randy Cogill, and Sanjay Lall.
\newblock Convexity of optimal control over networks with delays and arbitrary
  topology.
\newblock \emph{International Journal of Systems, Control and Communications},
  2\penalty0 (1-3):\penalty0 30--54, 2010.

\bibitem[Lessard and Lall(2011)]{lessard2011quadratic}
Laurent Lessard and Sanjay Lall.
\newblock Quadratic invariance is necessary and sufficient for convexity.
\newblock In \emph{Proceedings of the 2011 American Control Conference}, pages
  5360--5362. IEEE, 2011.

\bibitem[Bamieh and Voulgaris(2005)]{bamieh2005convex}
Bassam Bamieh and Petros~G Voulgaris.
\newblock A convex characterization of distributed control problems in
  spatially invariant systems with communication constraints.
\newblock \emph{Systems \& control letters}, 54\penalty0 (6):\penalty0
  575--583, 2005.

\bibitem[Mahajan et~al.(2012)Mahajan, Martins, Rotkowitz, and
  Y{\"u}ksel]{mahajan2012information}
Aditya Mahajan, Nuno~C Martins, Michael~C Rotkowitz, and Serdar Y{\"u}ksel.
\newblock Information structures in optimal decentralized control.
\newblock In \emph{2012 IEEE 51st IEEE Conference on Decision and Control
  (CDC)}, pages 1291--1306. IEEE, 2012.

\bibitem[Tolstaya et~al.(2020)Tolstaya, Gama, Paulos, Pappas, Kumar, and
  Ribeiro]{tolstaya2020learning}
Ekaterina Tolstaya, Fernando Gama, James Paulos, George Pappas, Vijay Kumar,
  and Alejandro Ribeiro.
\newblock Learning decentralized controllers for robot swarms with graph neural
  networks.
\newblock In \emph{Conference on Robot Learning}, pages 671--682. PMLR, 2020.

\bibitem[Gama et~al.(2020{\natexlab{a}})Gama, Tolstaya, and
  Ribeiro]{gama2020graph}
Fernando Gama, Ekaterina Tolstaya, and Alejandro Ribeiro.
\newblock Graph neural networks for decentralized controllers.
\newblock \emph{arXiv preprint arXiv:2003.10280}, 2020{\natexlab{a}}.

\bibitem[Khan et~al.(2020)Khan, Tolstaya, Ribeiro, and Kumar]{khan2020graph}
Arbaaz Khan, Ekaterina Tolstaya, Alejandro Ribeiro, and Vijay Kumar.
\newblock Graph policy gradients for large scale robot control.
\newblock In \emph{Conference on Robot Learning}, pages 823--834. PMLR, 2020.

\bibitem[Gama and Sojoudi(2021)]{gama2021distributed}
Fernando Gama and Somayeh Sojoudi.
\newblock Distributed linear-quadratic control with graph neural networks.
\newblock \emph{arXiv preprint arXiv:2103.08417}, 2021.

\bibitem[Gama et~al.(2020{\natexlab{b}})Gama, Bruna, and
  Ribeiro]{gama2020stability}
Fernando Gama, Joan Bruna, and Alejandro Ribeiro.
\newblock Stability properties of graph neural networks.
\newblock \emph{IEEE Transactions on Signal Processing}, 68:\penalty0
  5680--5695, 2020{\natexlab{b}}.

\bibitem[Bruna et~al.(2013)Bruna, Zaremba, Szlam, and LeCun]{bruna2013spectral}
Joan Bruna, Wojciech Zaremba, Arthur Szlam, and Yann LeCun.
\newblock Spectral networks and locally connected networks on graphs.
\newblock \emph{arXiv preprint arXiv:1312.6203}, 2013.

\bibitem[Kipf and Welling(2016)]{kipf2016semi}
Thomas~N Kipf and Max Welling.
\newblock Semi-supervised classification with graph convolutional networks.
\newblock \emph{arXiv preprint arXiv:1609.02907}, 2016.

\bibitem[Atwood and Towsley(2015)]{atwood2015diffusion}
James Atwood and Don Towsley.
\newblock Diffusion-convolutional neural networks.
\newblock \emph{arXiv preprint arXiv:1511.02136}, 2015.

\bibitem[Gama et~al.(2018)Gama, Marques, Leus, and
  Ribeiro]{gama2018convolutional}
Fernando Gama, Antonio~G Marques, Geert Leus, and Alejandro Ribeiro.
\newblock Convolutional neural network architectures for signals supported on
  graphs.
\newblock \emph{IEEE Transactions on Signal Processing}, 67\penalty0
  (4):\penalty0 1034--1049, 2018.

\bibitem[Goodfellow et~al.(2016)Goodfellow, Bengio, Courville, and
  Bengio]{goodfellow2016deep}
Ian Goodfellow, Yoshua Bengio, Aaron Courville, and Yoshua Bengio.
\newblock \emph{Deep learning}, volume~1.
\newblock MIT press Cambridge, 2016.

\bibitem[Tibshirani(1996)]{tibshirani1996regression}
Robert Tibshirani.
\newblock Regression shrinkage and selection via the lasso.
\newblock \emph{Journal of the Royal Statistical Society: Series B
  (Methodological)}, 58\penalty0 (1):\penalty0 267--288, 1996.

\bibitem[Donoho(2006)]{donoho2006compressed}
David~L Donoho.
\newblock Compressed sensing.
\newblock \emph{IEEE Transactions on information theory}, 52\penalty0
  (4):\penalty0 1289--1306, 2006.

\bibitem[Kingma and Ba(2014)]{kingma2014adam}
Diederik~P Kingma and Jimmy Ba.
\newblock Adam: A method for stochastic optimization.
\newblock \emph{arXiv preprint arXiv:1412.6980}, 2014.

\bibitem[Tu et~al.(2021)Tu, Robey, and Matni]{tu2021closing}
Stephen Tu, Alexander Robey, and Nikolai Matni.
\newblock Closing the closed-loop distribution shift in safe imitation
  learning.
\newblock \emph{arXiv preprint arXiv:2102.09161}, 2021.

\bibitem[Revay et~al.(2020)Revay, Wang, and Manchester]{revay2020convex}
Max Revay, Ruigang Wang, and Ian~R Manchester.
\newblock Convex sets of robust recurrent neural networks.
\newblock \emph{arXiv preprint arXiv:2004.05290}, 2020.

\end{thebibliography}
